\documentclass[twocolumn]{aastex631}
\usepackage{graphicx}
\usepackage{amsmath}
\usepackage{hyperref}
\usepackage{booktabs}

\newcommand{\ttm}[1]{#1}
\begin{document}
\title{Revisiting the Hubble Tension with DESI DR2 Baryon Acoustic Oscillation Observations and Machine Learning Methods}

\author[0000-0002-7003-7613]{Chenfa Zheng}
\affiliation{Institute for Frontiers in Astronomy and Astrophysics, Beijing Normal University, Beijing 102206, People's Republic of China}
\affiliation{School of Physics and Astronomy, Beijing Normal University, Beijing 100875, People's Republic of China}

\author[0000-0002-8870-981X]{Shuo Cao}
\affiliation{Institute for Frontiers in Astronomy and Astrophysics, Beijing Normal University, Beijing 102206, People's Republic of China}
\affiliation{School of Physics and Astronomy, Beijing Normal University, Beijing 100875, People's Republic of China}
\correspondingauthor{Shuo Cao}
\email{caoshuo@bnu.edu.cn}

\author[0009-0002-8489-1199]{Wuzheng Guo}
\affiliation{Institute for Frontiers in Astronomy and Astrophysics, Beijing Normal University, Beijing 102206, People's Republic of China}
\affiliation{School of Physics and Astronomy, Beijing Normal University, Beijing 100875, People's Republic of China}
\correspondingauthor{Wuzheng Guo}
\email{202331160017@mail.bnu.edu.cn}

\author[0009-0003-8646-573X]{Qiumin Wang}
\affiliation{Institute for Frontiers in Astronomy and Astrophysics, Beijing Normal University, Beijing 102206, People's Republic of China}
\affiliation{School of Physics and Astronomy, Beijing Normal University, Beijing 100875, People's Republic of China}

\author[0009-0001-4193-5684]{Xinyue Jiang}
\affiliation{Institute for Frontiers in Astronomy and Astrophysics, Beijing Normal University, Beijing 102206, People's Republic of China}
\affiliation{School of Physics and Astronomy, Beijing Normal University, Beijing 100875, People's Republic of China}
\affiliation{National Centre for Nuclear Research, Pasteura 7, PL-02-093 Warsaw, Poland}

\author[0000-0003-1308-7304]{Marek Biesiada}
\affiliation{National Centre for Nuclear Research, Pasteura 7, PL-02-093 Warsaw, Poland}
\correspondingauthor{Marek Biesiada}
\email{Marek.Biesiada@ncbj.gov.pl}

\author[0000-0002-6717-8810]{Tonghua Liu}
\affiliation{School of Physics and Optoelectronic, Yangtze University, Jingzhou 434023, People's Republic of China}

\received{2025 September 28}
\revised{2026 June 8}
\accepted{2026 June 11}
\published{2026 August 3}

\begin{abstract}
One of the key unresolved questions in modern cosmology is the Hubble tension, which arises from the inconsistency between determinations of the Hubble constant ($H_0$) based on nearby observations and those predicted by early-Universe data under the standard $\Lambda$CDM paradigm. Recent advances have identified the new Dark Energy Spectroscopic Instrument (DESI) survey as a promising observational tool for probing late-time cosmology. In this paper, we use the latest DESI DR2 baryon acoustic oscillation (BAO) measurements, combined with nonparametrically reconstructed Type Ia supernovae and cosmic chronometer data, to constrain the Hubble constant using three complementary machine learning methodologies: Gaussian process regression (GPR), artificial neural networks (ANNs), and long short-term memory (LSTM) networks. \ttm{We perform individual and joint constraints on $H_0$ at the effective redshifts of DESI DR2 tracers, yielding $H_0 = 61.8 \pm 5.3\,\mathrm{km\,s^{-1}\,Mpc^{-1}}$ (GPR), $66.3 \pm 5.2\,\mathrm{km\,s^{-1}\,Mpc^{-1}}$ (ANN), and $68.7 \pm 6.0\,\mathrm{km\,s^{-1}\,Mpc^{-1}}$ (LSTM). Our results exhibit excellent agreement with the recent Planck measurements within $1\sigma$, which highlight the potential of data-driven, model-independent methods to address the Hubble tension.} Finally, we further test the sensitivity of our results to potential systematics in the DESI data, emphasizing the importance of understanding systematic uncertainties in BAO surveys.
\end{abstract}


\section{Introduction}

A key objective of modern cosmology is to explore the structure of the Universe, based on the most widely accepted paradigm for describing its present state and future evolution. The Hubble constant ($H_0$), defined as the current expansion rate of the Universe, is a cornerstone of this endeavor. Precision measurements of $H_0$ are essential for validating the cosmological model and understanding the dynamic evolution of its fundamental components. As the keystone of the standard model, the precise determination of $H_0$ serves as a direct test of its validity. However, a profound discrepancy now threatens this paradigm: early-Universe constraints derived from the cosmic microwave background (CMB) via Planck satellite data ($H_0 = 67.4 \pm 0.5\ \mathrm{km\,s^{-1}\,Mpc^{-1}}$) \citep{Planck2020} significantly diverge from late-Universe estimates utilizing Cepheid-calibrated Type Ia supernovae (SN Ia) by the SH0ES collaboration ($H_0 = 73.04 \pm 1.04\ \mathrm{km\,s^{-1}\,Mpc^{-1}}$) \citep{Pantheon2022}, reaching a $>5\sigma$ tension. This ``Hubble tension'' has ignited a crisis in cosmology, fundamentally challenging the completeness of the $\Lambda$CDM model.

Concerted efforts have sought to measure $H_0$ through diverse pathways \citep{Rigault2015, Spergal2015, Vagnozzi2020, LiuYang2023}. Alternative probes such as strong-lensing time delays (H0LiCOW) \citep{suyu2017h0licow,2021MNRAS.503.2179Q,2022ApJ...939...37L}, compact radio quasars \citep{cao2017ultra,cao2017measuring,2022A&A...668A..51L}, and gravitational-wave standard sirens \citep{abbott2016observation,2020IJMPD..2950105Z,2024ApJ...965L..11L} offer intriguing cross-checks, though their current precision remains constrained by relatively sparse datasets \citep{2024SSRv..220...48B, 2017Natur.551...85A}. On the theoretical front, extensions to $\Lambda$CDM---including early dark energy to modulate the sound horizon or phantom energy to amplify late-time expansion---have shown partial success, yet they frequently face challenges when confronted with complementary cosmological observations \citep{Farrar2014, Valentino2017, Yang2018, Starobinsky2007, Bengochea2011}. The persistence of this tension underscores the critical need for fundamentally new approaches that reduce model dependence, coupled with larger, higher-precision datasets. To alleviate potential model-dependent biases, nonparametric reconstruction techniques have gained significant traction. For instance, \citet{keeley2021reconstructing} successfully employed Gaussian process regression (GPR) to test the mutual consistency between Pantheon supernovae and earlier SDSS/eBOSS BAO datasets, achieving a purely data-driven reconstruction of the expansion history. Further cementing the power of such model-independent frameworks, recent studies have demonstrated that nonparametric iterative smoothing techniques can robustly constrain key combinations like $H_0 r_d$ and cosmic curvature without assuming any underlying dark energy phenomenology \citep{l2025litmus}. Along similar lines, \citet{Bernal2016} demonstrated cosmic expansion consistency via joint BAO and SN Ia analyses, inferring $H_0$ through the parameter $r_{\rm d}$. However, this approach inherently reintroduces dependencies on early-Universe physics, which the Hubble tension itself brings into question. A transformative advance was achieved by \citet{Renzi2023}, who leveraged the distance duality relation (DDR) to derive $H_0$ solely from geometric observables, successfully circumventing rigid assumptions regarding dark energy or sound horizon scales.

BAOs originate from sound waves propagating in the primordial plasma of the early Universe. These waves froze at the epoch of recombination ($z \sim 1100$), imprinting a characteristic scale---the sound horizon $r_{\rm d}$---on the spatial distribution of matter. As the Universe expands, this scale stretches proportionally to cosmic distances, establishing BAO as a robust standard ruler for probing cosmic geometry \citep{2019MNRAS.488.3844R,2025ApJ...978L..33G}. The detection of BAO in galaxy surveys relies on measuring a distinct peak in the two-point correlation function at a comoving separation of $r_{\rm d} \sim 150\ \mathrm{Mpc}$, a relic of primordial density fluctuations. Pioneered by the Sloan Digital Sky Survey (SDSS) in 2005, BAO measurements have evolved considerably over the past two decades \citep{smee2013multi,percival2007measuring,dawson2012baryon}. Building upon SDSS’s ground-breaking work, the Dark Energy Spectroscopic Instrument (DESI) has substantially expanded both the depth and precision of BAO analyses. Recently, the second data release (DR2) of DESI \citep{DESI2025} has demonstrated remarkable advancements in sky coverage and data quality compared to DR1. The sample sizes of emission-line galaxies (ELGs), luminous red galaxies (LRGs), and quasars (QSOs) have expanded by factors of 2.7, 2.1, and 1.7, respectively, accompanied by a completeness improvement in ELG assignment from 35.2\% to 53.7\%. The surveyed sky area has increased by 54\% for QSOs and 75\% for LRGs/ELGs, while the Bright Galaxy Survey footprint has grown by 65\%. These improvements stem from strategic observational refinements, which simultaneously reduced priority-masked regions and enhanced effective spatial sampling. Furthermore, the DR2 quasar catalog has doubled in size (reaching 820,000 quasars at $z > 2.09$), with 67\% of targets observed at least twice, thereby significantly improving the signal-to-noise ratio of $\mathrm{Ly}\alpha$ forest measurements. Together, these enhancements boost DR2's statistical constraining power by 30\%--50\% over DR1, enabling unprecedented precision in cosmological parameter estimation. Consequently, this updated data release yields highly anticipated BAO results, providing independent geometric measurements through the transverse comoving distance $D_{\rm M}$ and the line-of-sight comoving distance $D_{\rm H}$. This multitracer framework effectively mitigates potential biases and paves the way for tighter cosmological constraints. In fact, recent extended analyses of DESI DR2 BAO measurements utilizing a broad range of nonparametric approaches (including Gaussian processes) have uncovered robust hints of dynamical dark energy at low redshifts ($z \lesssim 0.3$), underscoring the necessity of flexible, data-driven frameworks to capture potential deviations from standard $\Lambda$CDM \citep{lodha2025extended}.

Building upon these methodological and observational advances, the next-generation constraint on $H_0$ can be robustly performed by integrating the latest DESI DR2 BAO data---which span a wide redshift range with subpercent statistical precision---into the model-independent DDR framework. This synthesis not only tests the redshift-consistency of $H_0$ but also probes potential systematics across cosmic epochs, offering new pathways to disentangle observational biases from true cosmological anomalies. The structure of this paper is organized as follows. In Section~\ref{Sec:2}, we establish the theoretical framework and introduce the observational data used in this analysis. Section~\ref{Sec:3} describes the implementation and optimization details of our nonparametric reconstruction techniques, specifically GPR, artificial neural networks (ANNs), and long short-term memory networks (LSTM). In Section~\ref{Sec:4}, we present a cross-method comparative analysis of the derived Hubble constant, followed by a comprehensive discussion on the implications for the Hubble tension.

\section{Method and Observations}\label{Sec:2}

The Hubble constant can be independently estimated by combining the reconstructed cosmic chronometer function $H(z)$, the luminosity distance from SN Ia, and the angular diameter distance from BAO measurements. Relying on the DDR,
\begin{equation}
    \frac{d_{\rm L}(z)}{(1+z)^2 d_{\rm A}(z)} = 1,
    \label{Eq:DDR}
\end{equation}
which holds under the assumptions of photon number conservation and null geodesic propagation, we can formulate the Hubble constant in terms of three independent observables:
\begin{equation}
    H_0 = \frac{1}{(1+z)^2} \frac{\Xi_{\rm SN Ia}(z)}{\Xi_{\rm BAO}(z)} H_{\rm CC}(z),
    \label{Eq:H0}
\end{equation}
where $ \Xi_{\rm SN Ia}(z) = H_0 d_{\rm L}(z) $ is derived from SN Ia apparent magnitudes (unanchored distance), $ \Xi_{\rm BAO}(z) = H(z)d_{\rm A}(z) $ is obtained from the DESI DR2 BAO measurements, and $ H_{\rm CC}(z) $ is reconstructed at the corresponding redshift using CC data. In this work, six high-precision BAO measurements from the DESI DR2 sample, spanning the redshift range $ 0.510 \le z \le 2.330 $, are utilized. To ensure consistency, both SN Ia and CC datasets are interpolated to these redshifts using nonparametric reconstruction techniques, preserving the global trends and error correlations across redshifts while ensuring a stable estimation of $ H_0 $ with minimal model bias. 

\subsection{DESI DR2 BAO Observations}

In our analysis, we employ the latest BAO measurements from the DESI Data Release 2 (DR2) \citep{DESI2025}, which provide six high-precision data points across the redshift range $ z \in [0.510, 2.330] $. These data points are derived from LRGs, ELGs, quasars (QSO), and Ly$\alpha$ forest (Ly$\alpha$) tracers. 
Notably, the DR2 data show improved parameter precision at identical redshift points compared to DR1, while additionally introducing new measurements at the QSO ($z=1.484$) regime. The DESI DR2 has significantly extended the redshift coverage beyond BOSS, with the number of objects observed in these releases far exceeding the BOSS sample size, representing a significant improvement in BAO precision. 

The BAO signal arises from sound waves propagating in the early Universe and imprinted as a preferred clustering scale in the distribution of galaxies. The DESI survey measures both the transverse and radial components of this scale, providing access to the angular diameter distance $d_{\rm A}(z)$ and the Hubble parameter $H(z)$. However, instead of relying on model-dependent distances such as $ D_{\rm M} $ or the sound horizon scale $ r_{\rm d} $, we construct the observable combination
\begin{equation}
    H(z)d_{\rm A}(z) = \frac{c}{1+z} \cdot \frac{D_{\rm M}(z)}{D_{\rm H}(z)},
    \label{Eq:HzdA}
\end{equation}
which can be directly obtained from the BAO angular and radial measurements and used as the denominator in our model-independent expression for the Hubble constant in Equation~(\ref{Eq:H0}). Due to the high precision of the DESI BAO data in DR1 and DR2—reaching subpercent accuracy in some redshift bins—these datasets have already been widely adopted in recent cosmological studies \citep{Cort2024, Giar2024, Wang2024}, including investigations into dark energy, modified gravity, and the Hubble tension \citep{Bousis2024, Jia2024, WangHu2024, silva2025new}. 

\subsection{Cosmic Chronometer and SN Ia Observations}

The Hubble parameter $ H(z) $ is also independently measured via cosmic chronometers (CC), which use the differential aging of massive, passively evolving galaxies to determine the expansion rate of the Universe
\begin{equation}
    H(z) = -\frac{1}{1+z} \frac{{\rm d}z}{{\rm d}t}.
    \label{Eq:Hzfromdt}
\end{equation}
Massive early-type galaxies with stellar masses $ M_{\rm stars} > 10^{11} M_\odot $ are ideal cosmic clocks due to their old stellar populations and well-understood evolution. By measuring $ {\rm d}t $ across redshift intervals $ {\rm d}z $, one obtains a direct estimate of $H(z)$ that relies on stellar astrophysics rather than assuming a background cosmological model. In this study, we use the most updated compilation of 32 CC measurements as presented in \citet{Qi2023}, spanning the redshift range $ z = 0.07 $ to $ z = 1.965 $. \ttm{These data points include both statistical and systematic uncertainties associated with stellar population modeling and spectroscopic age determinations, and the full covariance matrix can be retrieved from \texttt{CCcovariance}\footnote{\url{https://gitlab.com/mmoresco/CCcovariance}}\citep{moresco2012improved,moresco2015raising,Moresco2016}.} To apply Equation~(\ref{Eq:H0}), we require estimates of $ H(z) $ at the exact redshifts of the DESI DR2 BAO data points, which is achieved by performing nonparametric reconstruction of the CC dataset using GPR, ANN, and LSTM. These methods yield smooth expansion histories consistent with the predictions of the $\Lambda$CDM model, allowing us to evaluate Equation~(\ref{Eq:H0}) at each of the DESI BAO redshifts. We note that the reconstructed $H(z)$ at the Ly$\alpha$ point, with an effective redshift of $z=2.330$, requires extrapolation since it lies beyond the maximum redshift of the CC data ($z \le 1.965$).

We incorporate SN Ia observations to obtain the quantity $ H_0 d_{\rm L}(z) $, which plays a crucial role in our model-independent estimation of the Hubble constant. This combination can be directly derived from the observed apparent magnitudes of SN Ia, without requiring any external information from the CMB. The key relation is given by
\begin{equation}
    m_{\rm B} = 5\log_{10}(H_0 d_{\rm L}) - 5a_{\rm B},
    \label{Eq:mb}
\end{equation}
where $ m_{\rm B} $ is the standardized apparent magnitude of the SN Ia at peak brightness, and $ a_{\rm B} $ is the intercept of the magnitude-redshift relation, which absorbs both the absolute magnitude $M_B$ and the Hubble constant (specifically, $a_B = -0.2M_B + \log_{10} H_0 - 5$). Rearranging Equation~(\ref{Eq:mb}), we obtain the unanchored distance:
\begin{equation}
    H_0 d_{\rm L}(z) = 10^{0.2 m_{\rm B} + a_{\rm B}}.
    \label{Eq:H0dL}
\end{equation}
This formulation bypasses the need to independently calibrate the absolute magnitude $ M_{\rm B} $, and therefore avoids associated calibration uncertainties. Instead, we fix the value of the nuisance parameter $ a_{\rm B} = 0.71273 \pm 0.00176 $ as inferred directly from SN Ia observations in the low-redshift Hubble flow \citep{Riess_2016}. This value has been shown to be consistent across different assumptions about the expansion history, as reviewed in the latest SH0ES analysis \citep{Pantheon2022}, and does not introduce significant model dependence. The SN Ia data used in this work come from the Pantheon+ sample \citep{Pantheon2022,scolnic2022pantheon+}, which provides an updated compilation of over 1700 SN Ia light curves spanning a redshift range from $ z \sim 0.001 $ to $ z \sim 2.26 $. The catalog provides standardized light-curve parameters and peak magnitudes $ m_{\rm B} $, derived using a consistent methodology across different surveys and instruments. These standardized measurements have been rigorously reprocessed to account for selection effects, intrinsic scatter, and systematic biases.

\section{Machine Learning Methods}\label{Sec:3}

A brief overview of the nonparametric reconstruction techniques used in this work, including GPR, ANNs, and LSTM networks, is provided below.

\subsection{Gaussian Process Regression}\label{Sec:2-GPR}

The GPR method is implemented using the \texttt{GaPP} Python code, which is widely used in cosmology studies \citep{Seikel2012a,Seikel2012b,2020ApJ...888..113W,liu2021revisiting,Zhang:2018gjb, zhang2025H}. In this framework, the reconstructed function $f(z)$ is assumed to follow a Gaussian process (GP), where the correlation between its values at two redshifts $z$ and $\tilde{z}$ is determined by a covariance function $k(z,\tilde{z})$. Following foundational GPR cosmography frameworks \citep{shafieloo2012gaussian}, we specifically employ the squared exponential kernel defined as 
\begin{equation}
    k(z, \tilde{z}) = \sigma_f^2\exp \left[-\frac{(z-\tilde{z})^2}{2l^2}\right].
\end{equation} This function depends on two hyperparameters: $\sigma_f$, which characterizes the typical amplitude of variations in the data, and $\ell$, which denotes the characteristic length scale over which significant changes in $f(z)$ occur.

The regression incorporates a mean function that serves to create an initial estimate of the function prior to fitting the data. This mean function can be informed by a theoretical cosmological model, allowing the GPR to leverage existing knowledge and potentially achieve a more accurate reconstruction. However, as pointed out by previous studies \citep{shafieloo2012gaussian,Hwang2023}, simply optimizing the hyperparameters with a fixed mean function cannot completely eliminate user-dependent biases, as the reconstruction may retain a ``memory'' of the initial choice and remain anchored to the chosen theoretical model. To ensure a strictly model-independent reconstruction, we adopt a ``full-marginalization'' approach by jointly marginalizing over the GPR hyperparameters and the CPL parameterization \citep{chevallier2001accelerating,linder2003exploring} via Markov Chain Monte Carlo (MCMC) sampling, thus effectively removing the dependence on the initial cosmological assumptions. In this Bayesian framework, we adopt uniform priors for both the physical parameters of the fiducial mean function and the GPR hyperparameters to prevent any user-dependent anchoring. For the CC data, the prior ranges are conservatively set as follows: $H_0 \in [50, 100] \, \mathrm{km\,s^{-1}\,Mpc^{-1}}$, $\Omega_m \in [0, 1]$, $w_0 \in [-2, 0]$, $w_a \in [-2, 2]$, $\sigma_f \in [0, 200]$, and the length scale $l \in [0.01, 10]$. For the SN Ia data, since the reconstructed observable is the unanchored luminosity distance $H_0 d_L(z)$ which spans a significantly larger numerical scale, the prior for the amplitude hyperparameter is expanded to $\sigma_f \in [0, 10^7]$. Notably, the $H_0$ parameter is naturally factored out in the SN Ia mean function, reducing the parameter space to $(\Omega_m, w_0, w_a, \sigma_f, l)$. The sampling is performed using the Affine Invariant MCMC Ensemble sampler (\texttt{emcee}\footnote{\url{https://github.com/dfm/emcee}}) \citep{emcee}. \ttm{An identical MCMC configuration is adopted for both datasets to achieve a standardized, consistent, and robust reconstruction framework. Considering the high dimensionality and complexity of the parameter space, each walker is evolved for a total of 15,000 steps to enable exhaustive sampling of the full parameter volume. The sampling efficiency is well confirmed by well-behaved mean acceptance fractions, which typically fall within the MCMC-recommended optimal range of $0.2$–$0.5$. Furthermore, to rigorously assess the convergence of our MCMC sampling, we compute the integrated autocorrelation time $\tau$ for each parameter using the built-in diagnostics of the \texttt{emcee} toolkit. All MCMC chains are verified to exceed the standard convergence threshold of $50\tau$. Finally, we dynamically discard the initial $5\max(\tau)$ steps as the burn-in phase and thin the remaining chains based on the autocorrelation time. This rigorous sampling treatment yields an effective sample size of more than $\mathcal{O}(10^3)$ independent samples, ensuring stable, high-precision posterior parameter inference for our subsequent analysis.}

\begin{figure*}[htbp]
    \centering
    \includegraphics[width=0.7\linewidth]{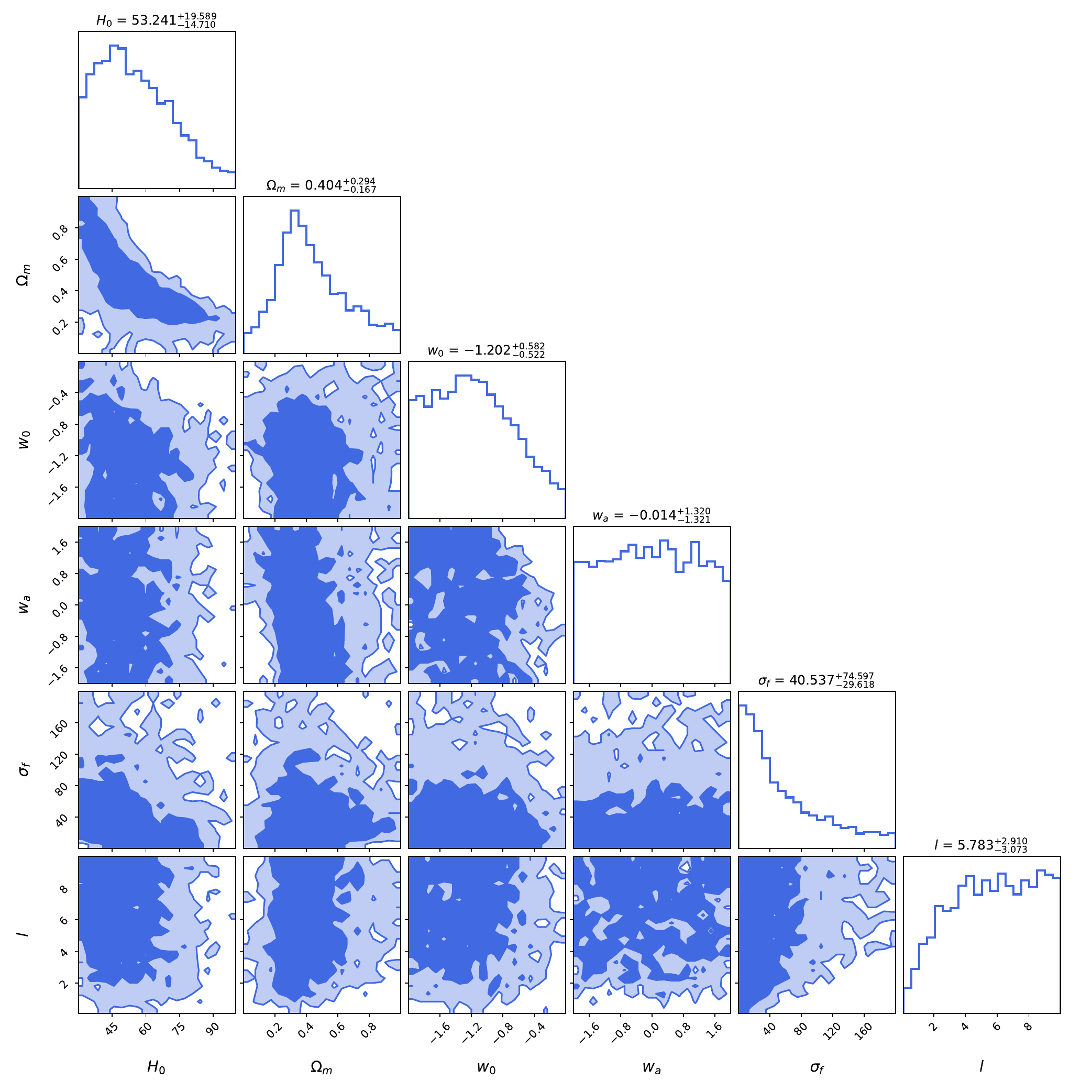}
    \caption{\ttm{The 1D and 2D marginalized posterior distributions for $(H_0, \Omega_m, w_0, w_a, \sigma_f, l)$ in the CC reconstruction with GPR.}}
    \label{fig:mcmc_corner_cc}
\end{figure*}

\begin{figure*}[htbp]
    \centering
    \includegraphics[width=0.7\linewidth]{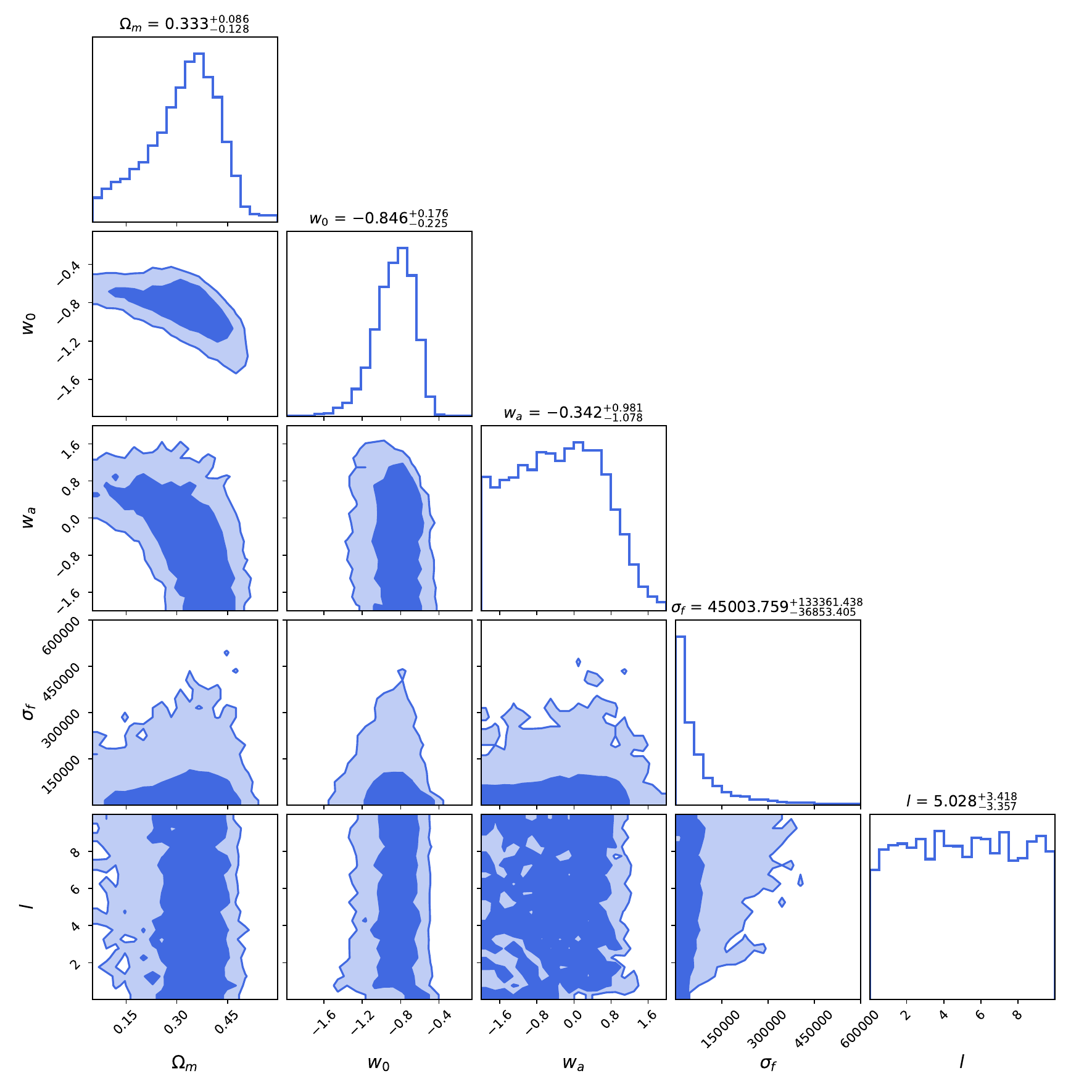}
    \caption{\ttm{The 1D and 2D marginalized posterior distributions for $(\Omega_m, w_0, w_a, \sigma_f, l)$  in the SN Ia reconstruction with GPR.}}
    \label{fig:mcmc_corner_snia}
\end{figure*}

Given the significant computational overhead of GPR on the large Pantheon+ dataset ($N \sim 1300$), we implement several numerical optimizations to ensure feasibility. Specifically, we precompute the squared-distance matrix for the kernel function and utilize highly optimized, vectorized cumulative integration for the CPL mean function. This allows for $\mathcal{O}(1)$ interpolation of the theoretical $H_0 d_L$ during the MCMC walk, effectively bypassing expensive numerical integration at each step. Furthermore, we calculate the log-likelihood using Cholesky decomposition instead of direct matrix inversion to enhance numerical stability and execution speed. Figures~\ref{fig:mcmc_corner_cc} and \ref{fig:mcmc_corner_snia} present the marginalized posterior distributions. The well-constrained, unimodal posteriors confirm that the full marginalization successfully captures the intrinsic uncertainty of both the mean function and the covariance structure. These posteriors are subsequently used to draw 1000 predictive multivariate Gaussian samples for the final cosmological reconstructions.

\subsection{Artificial Neural Networks}\label{Sec:2-ANN}

While the computational bottleneck of GPR's matrix operations can be significantly mitigated by modern GPU-accelerated linear algebra libraries, its scaling with extremely massive datasets remains a consideration. As an alternative, nonparametric methods based on ANNs rely on a feed-forward learning structure, offering highly efficient scaling. However, this computational flexibility comes at the cost of interpretability; unlike GPR, where the covariance kernel provides clear statistical meaning, neural networks inherently operate as "black boxes" with less transparent internal mechanics. To ensure full methodological transparency, the specific architecture, inputs, outputs, and training details of our network are explicitly defined as follows. The ANN is constructed as a fully connected feed-forward network. The sole input of the network is the redshift $z$, and the output is a two-dimensional vector containing the reconstructed theoretical function (e.g., $H(z)$ or $H_0 d_L(z)$) and its corresponding $1\sigma$ statistical uncertainty. Prior to the training process, both the input redshift and the target data are normalized using z-score scaling to accelerate convergence. For the internal architecture (corresponding to the \texttt{rec\_2} configuration in the \texttt{ReFANN}\footnote{\url{https://github.com/Guo-Jian-Wang/refann}} code), we adopt the exponential linear unit (ELU) as the activation function and implement batch normalization (BN) to ensure stable gradient propagation; dropout layers are explicitly excluded to maintain the fidelity of the variance estimation. The network is trained using the Adam optimizer aiming to minimize the L1 loss function (mean absolute error). During the training, the learning rate is set to decay exponentially from an initial value of $0.1$ down to $10^{-8}$. Regarding the stopping criterion, the training process is deterministically terminated after a fixed maximum of 30,000 iterations to ensure the full convergence of the loss function.

While the aforementioned training configurations are fixed throughout this work, the specific structural topology (i.e., the number of hidden layers and neurons) must be carefully tuned. To ensure an optimized network structure without suffering from overfitting or underfitting, the performance of ANN with various layers is benchmarked with their risk based on the bias--variance trade-off \citep{wasserman2001non}: Assuming the reconstructed function $f_\text{r}$ (from a certain neuron network configuration) successfully reproduces an estimate of unknown function $f$, one can evaluate its bias to the function $\mathcal{B}(z_i) = E[f_r(z_i)] - f(z_i)$ and its variance $\mathcal{V}(z_i) = E[(f_r(z_i) - E(f_r(z_i)))^2]$ at each mock data $z_i$. The overall magnitude of this bias across the entire mock dataset can be intuitively characterized by the Root Mean Square Error (RMSE), defined as
\begin{equation}
    \text{RMSE} = \sqrt{\frac{1}{N}\sum_{i}\left[\mathcal{B}(z_i)\right]^2},
\end{equation}
where $N$ is the total number of data points. Since RMSE provides a direct measure of the average systematic deviation with the same physical units as the target function, we will utilize it to represent the bias performance in our subsequent tabular comparisons. Then the performance of a specific configuration could be tested with its Risk, defined as
\begin{eqnarray}
    \text{Risk} & = &\sum_{i}\left[\mathcal{B}(z_i)\right]^2 + \sum_{i}\mathcal{V}(z_i)\nonumber\\
    & = & \text{Bias}^2 + \text{Variance},
\end{eqnarray}
where the sum denotes for all mock data for a certain fiducial $h$, represented by bias and variance, respectively. Therefore, minimizing the risk naturally balances the ability of the reconstruction reproducing the unknown function $f$ and its smoothness. Practically, four configurations of hidden layers are tested, namely, \text{C11}, one hidden layer with 4096 neurons; \text{C12}, one hidden layer with 8192 neurons; \text{C21}, two hidden layers with 2048 neurons each; and \text{C22}, two hidden layers with 4096 neurons each. These four configurations mimic the neuron network applied in this work (C11) and its minimal variation on number of hidden layers and/or neurons (C12, C21, and C22). 

\begin{figure*}[htbp]
    \centering
    \includegraphics[width=0.65\textwidth]{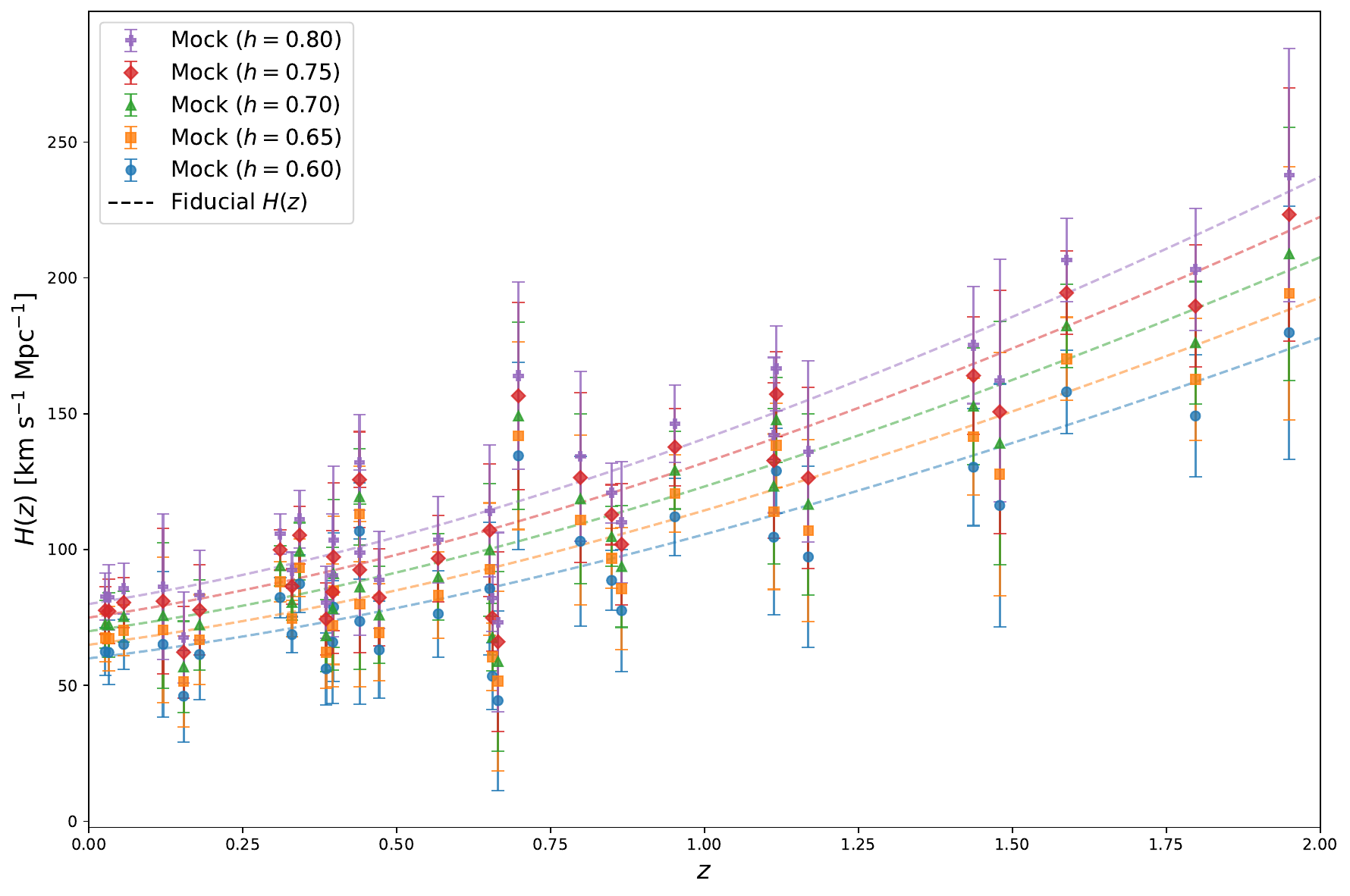} 
    \caption{The simulated CC datasets. The dashed lines represent the underlying fiducial flat $\Lambda$CDM models with different $H_0$ values, while the error bars illustrate the realistic redshift-dependent envelope infused with Gaussian noise.}
    \label{fig:mock_data}
\end{figure*}

To avoid overfitting on real $H(z)$ reconstruction during the benchmark, mock $H(z)$ data (as shown in Figure~\ref{fig:mock_data}) are applied to the training and testing process. In detail, the generation pipeline of the mock dataset is strictly designed to mimic the statistical properties and the redshift-dependent error distribution of the actual CC observations. The redshifts $z_i$ of the mock data are sampled directly from the redshift distribution of the observed CC sample, which ensures the simulated dataset preserves the genuine observational selection function. Then a fiducial flat $\Lambda$CDM model is adopted to compute the theoretical Hubble parameter $H_\text{th}(z_i)$ at each redshift, given by $H_\text{th}(z) = H_0 \sqrt{\Omega_m(1+z)^3 + (1-\Omega_m)}$, where $\Omega_m$ is fixed to $0.3$ and $H_0$ is set to a predetermined baseline value. To accurately reflect the deterioration of observational signal-to-noise ratio at higher redshifts, a symmetric linear error envelope is constructed from the real dataset. In detail, a central linear trend $L_0(z)$ is fitted to the observed uncertainties $\sigma_{H(z)}$. By specifying a lower boundary $L_1(z)$ based on the minimal observational scatter and computing its symmetric upper boundary $L_2(z)$ with respect to $L_0(z)$, a physically reasonable envelope is formed. The mock uncertainty $\sigma_{\text{mock}, i}$ for each point is then uniformly drawn from the interval $[\min(L_1, L_2), \max(L_1, L_2)]$. Finally the Gaussian random noise is injected into the fiducial value, and the mock Hubble parameter $H_{\text{mock}, i}$ strictly follows $H_{\text{mock}, i} \sim \mathcal{N}(H_\text{th}(z_i), \sigma_{\text{mock}, i}^2)$.

To thoroughly evaluate the network's stability against different background cosmologies, these mock data are further categorized into distinct configurations, where the test cases are generated using different fiducial Hubble constants, $H_{0,\mathrm{fid}} = h\cdot 100\ \mathrm{km\ s^{-1}\ Mpc^{-1}}$ (see Table \ref{tab:conf_suffix}). The nomenclature for these configurations appends a suffix to the network architecture denoting the dataset type and the corresponding $h$ value. Furthermore, to systematically present the reconstruction performance, Table \ref{tab:conf_suffix} reports four fundamental benchmarks: the reduced chi-square ($\chi^2$) to mock data, and the RMSE, variance, and risk to the corresponding fiducial function at redshift of mock data ($H_\text{th}(z_i)$). The $\chi^2$ metric evaluates the data-fitting capability, measuring how well the ANN prediction aligns with the noisy mock observations. Conversely, the risk, alongside the RMSE (representing the average systematic bias) and variance, serves as a quantitative indicator of the truth-recovery capability, representing the deviation of the reconstructed function from the underlying theoretical fiducial model.

\begin{table}[htbp]
  \centering
  \caption{Configuration notation, parameters, and benchmarks on test cases corresponding to different neuron network architectures in ANN reconstruction for CC data. The details of each column are as follows: 
  Column 1 (Notation): the specific network configuration and test case identifier; 
  Column 2 ($N$): the number of mock data points used for evaluation; 
  Column 3 ($h$): the fiducial dimensionless Hubble parameter; 
  Column 4 ($\chi^2$): the reduced chi-square indicating the goodness of fit to the mock data; 
  Column 5 (RMSE): the root mean square error representing the average systematic bias, in units of $\mathrm{km\,s^{-1}\,Mpc^{-1}}$; 
  Column 6 (Var.): the sum of the variance $\sum \mathcal{V}$ representing the model uncertainty, in units of $10^3\,\mathrm{km^2\,s^{-2}\,Mpc^{-2}}$; 
  Column 7 (Risk): the total risk balancing bias and variance, in units of $10^3\,\mathrm{km^2\,s^{-2}\,Mpc^{-2}}$.}
  \begin{tabular}{lcccccc}
    \hline\hline
    Notation & $N$ & $h$ & $\chi^2$ & RMSE & Var. & Risk \\
    \midrule
    C11CC60 & 32 & 0.60 & 0.83 & 3.52 & 12.0 & 12.4 \\
    C11CC65 & 32 & 0.65 & 0.83 & 3.86 & 12.1 & 12.5 \\
    C11CC70 & 32 & 0.70 & 0.83 & 3.57 & 12.0 & 12.4 \\
    C11CC75 & 32 & 0.75 & 0.83 & 3.89 & 12.1 & 12.6 \\
    C11CC80 & 32 & 0.80 & 0.83 & 3.60 & 12.0 & 12.4 \\
    \midrule
    C12CC60 & 32 & 0.60 & 0.83 & 3.82 & 12.1 & 12.5 \\
    C12CC65 & 32 & 0.65 & 0.83 & 3.83 & 11.9 & 12.4 \\
    C12CC70 & 32 & 0.70 & 0.83 & 3.88 & 11.9 & 12.4 \\
    C12CC75 & 32 & 0.75 & 0.83 & 3.56 & 12.1 & 12.5 \\
    C12CC80 & 32 & 0.80 & 0.83 & 3.93 & 12.1 & 12.6 \\
    \midrule
    C21CC60 & 32 & 0.60 & 0.86 & 4.47 & 12.0 & 12.6 \\
    C21CC65 & 32 & 0.65 & 0.84 & 4.09 & 12.2 & 12.8 \\
    C21CC70 & 32 & 0.70 & 0.87 & 4.52 & 11.9 & 12.6 \\
    C21CC75 & 32 & 0.75 & 0.86 & 4.53 & 12.0 & 12.7 \\
    C21CC80 & 32 & 0.80 & 0.83 & 4.27 & 12.1 & 12.7 \\
    \midrule
    C22CC60 & 32 & 0.60 & 0.88 & 4.37 & 12.4 & 13.0 \\
    C22CC65 & 32 & 0.65 & 0.87 & 4.45 & 12.1 & 12.7 \\
    C22CC70 & 32 & 0.70 & 0.88 & 4.34 & 12.0 & 12.6 \\
    C22CC75 & 32 & 0.75 & 0.88 & 4.41 & 12.2 & 12.9 \\
    C22CC80 & 32 & 0.80 & 0.82 & 4.62 & 12.1 & 12.8 \\
    \hline\hline
  \end{tabular}
  \label{tab:conf_suffix}
\end{table}

\begin{figure*}[htbp]
    \centering
    \begin{minipage}{0.45\textwidth}
        \centering
        \includegraphics[width=\linewidth]{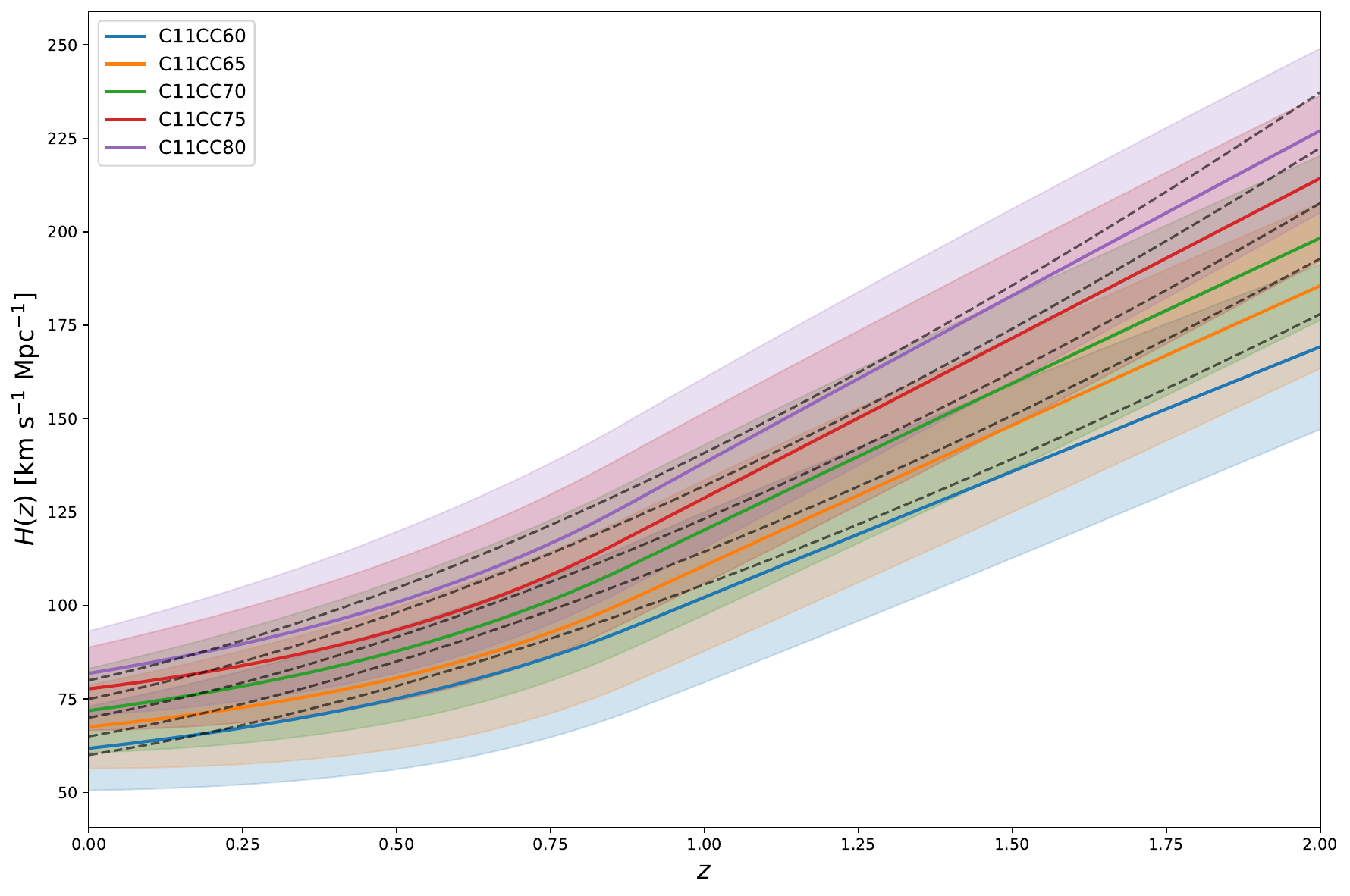}
        \centerline{(a) C11: 1 hidden layer, 4096 neurons}
    \end{minipage}\hfill
    \begin{minipage}{0.45\textwidth}
        \centering
        \includegraphics[width=\linewidth]{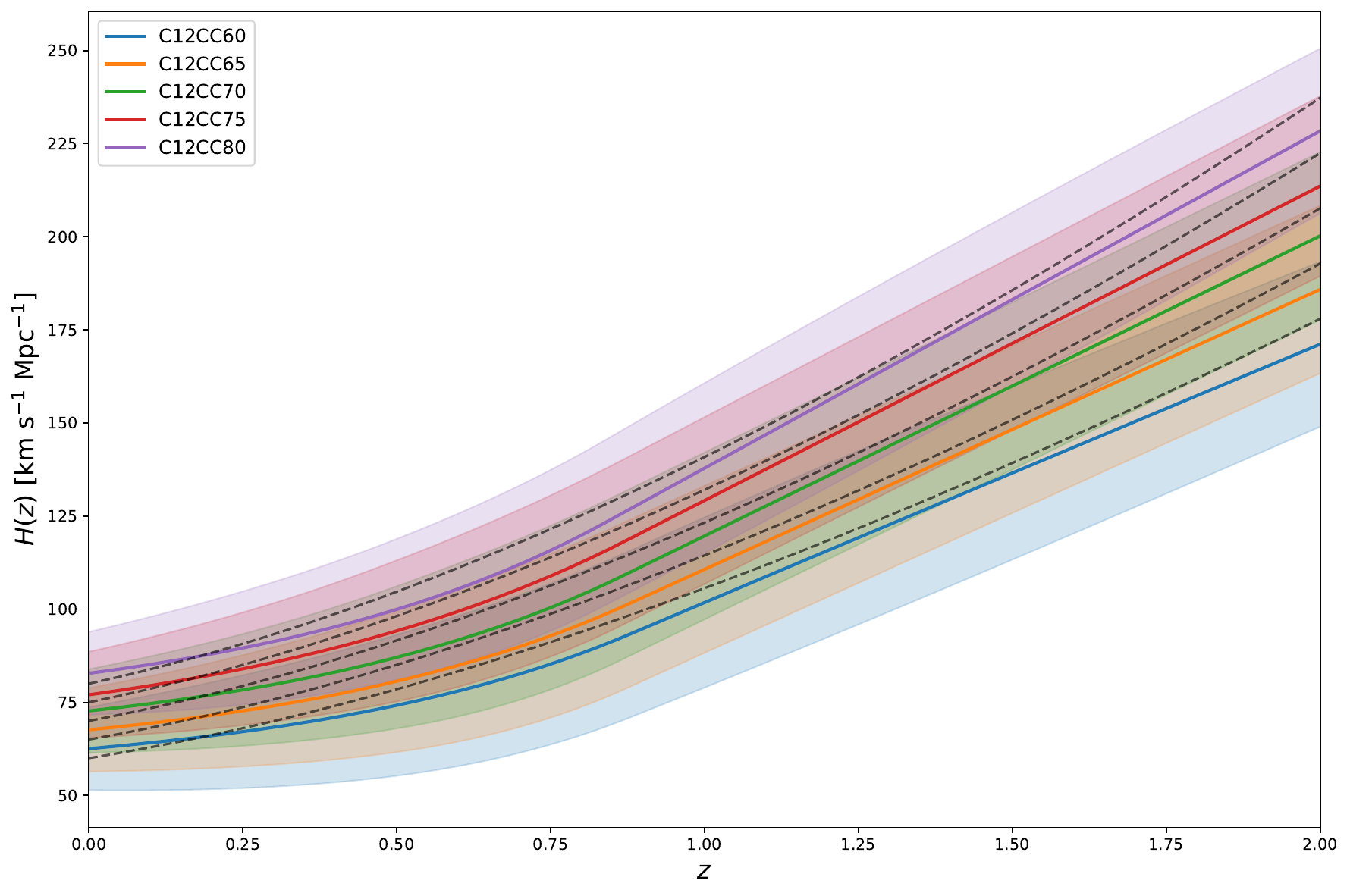}
        \centerline{(b) C12: 1 hidden layer, 8192 neurons}
    \end{minipage}
    
    \vspace{1.5em}
    
    \begin{minipage}{0.45\textwidth}
        \centering
        \includegraphics[width=\linewidth]{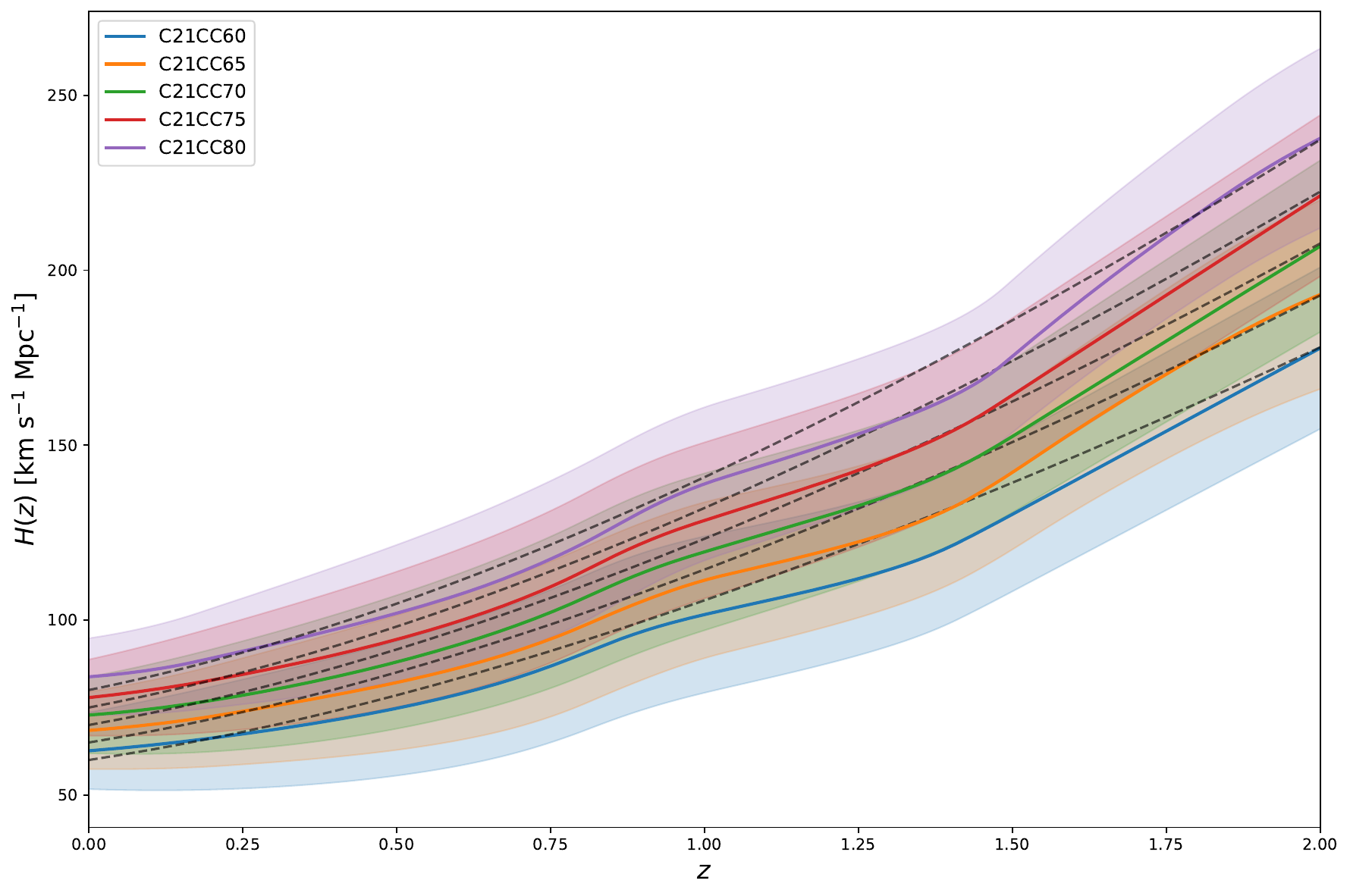}
        \centerline{(c) C21: 2 hidden layers, 2048 neurons}
    \end{minipage}\hfill
    \begin{minipage}{0.45\textwidth}
        \centering
        \includegraphics[width=\linewidth]{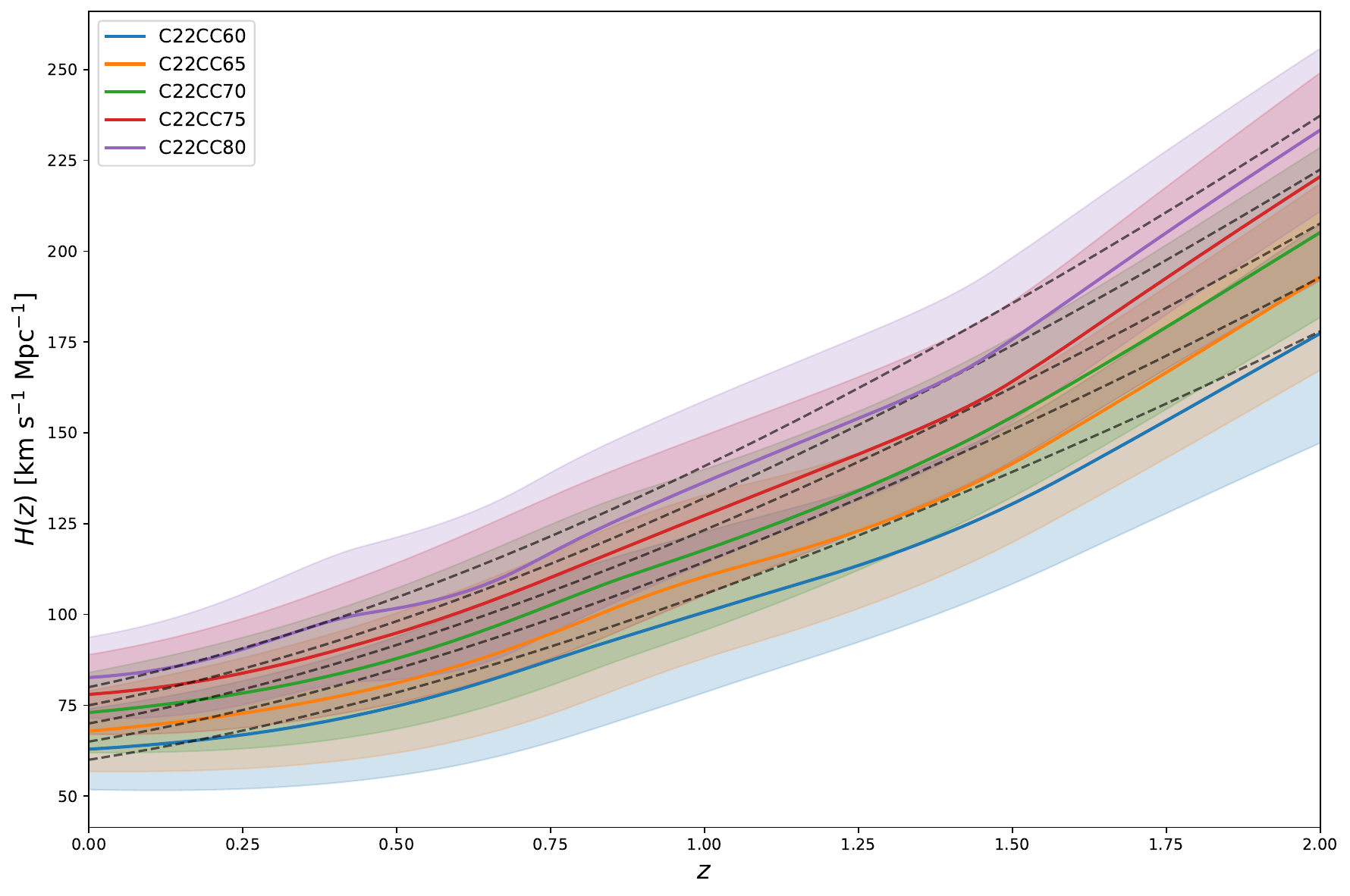}
        \centerline{(d) C22: 2 hidden layers, 4096 neurons}
    \end{minipage}
    
    \caption{The ANN reconstruction results applied to the mock CC datasets across four different network architectures. The single-layer configurations (panels (a) and (b)) demonstrate stable recovery of the fiducial cosmology, whereas the dual-layer configurations (panels (c) and (d)) exhibit severe overfitting behaviors, tightly winding around the noisy scattered points.}
    \label{fig:ann_results}
\end{figure*}

Furthermore, to rigorously assess the aforementioned configurations (\text{C11} to \text{C22}) and to isolate the intrinsic architectural performance from random statistical fluctuations, identical random seeds are utilized across comparative testings. By feeding these strictly paired mock datasets into the ANN, the aforementioned risks can be quantitatively evaluated, enabling the robust selection of an optimal network architecture capable of recovering the underlying physical laws without suffering from performance degradation. The visual results of the CC performance evaluation are illustrated in Figure \ref{fig:ann_results}. Visual inspection directly corroborates the quantitative benchmarks in Table \ref{tab:conf_suffix}: for the sparse CC data, while the single-layer configurations (C11 and C12) smoothly recover the fiducial model, the dual-layer architectures (C21 and C22) exhibit noticeable instability, struggling to effectively extract the underlying physical trend due to their excessive structural complexity.

Beyond visual inspection, the quantitative results summarized in Table \ref{tab:conf_suffix} vividly illustrate the intricate interplay between network architecture, dataset volume, and cosmological parameter stability. Primarily, the evaluation on the CC datasets confirms the absolute robustness of the single-layer configurations against different underlying cosmologies. For both C11 and C12 architectures, as the fiducial parameter $h$ varies broadly from $0.60$ to $0.80$, the reduced $\chi^2$ remains highly stable at $0.83$, and the reconstruction error (RMSE) is strictly confined within a narrow margin of $3.52 \sim 3.93 \ \mathrm{km\ s^{-1}\ Mpc^{-1}}$. This consistency provides strong statistical evidence that the ANN method acts as a genuinely model-independent reconstructor, successfully evading systematic biases induced by prior assumptions of the Hubble constant.

Moreover, the comparative analysis between single-layer and dual-layer architectures on the CC dataset highlights the limitations of deploying overly complex models on sparse observations. Unlike scenarios with massive datasets where deeper networks might achieve better fits, the dual-layer models (C21 and C22) here not only yield higher $\chi^2$ values ($0.84\sim 0.88$) but also exhibit elevated truth-recovery risks. The RMSE noticeably increases to $4.09\sim 4.62\ \mathrm{km\ s^{-1}\ Mpc^{-1}}$, demonstrating that excessively deep networks introduce greater systematic deviations when constrained by limited data (i.e., $N=32$). We note here that a more comprehensive comparison of different configurations on sparse datasets can be found in \cite{Wang2020}, which reveals exactly the same conclusion. These comprehensive mock tests underscore that employing a highly complex network topology is not inherently advantageous for cosmological reconstructions involving limited observations. To effectively suppress structural instability and reliably extract pure physical signals from the sparse CC dataset, the single-layer architecture with 4,096 neurons (C11) proves to be the most optimal and stable configuration, strictly justifying its adoption for the analysis in this work.

To systematically investigate the network performance on datasets characterized by massive volume and strong non-diagonal covariance, we extend the risk evaluation to the Pantheon+ SN Ia sample ($N=1580$ with the removal of calibration sample). As presented in Table \ref{tab:sn_risk}, exploring a broader parameter space---from single-layer (C1x) up to three-layer (C3x) configurations---reveals a U-shaped risk trend. The single-layer models (C11SN and C12SN) exhibit severe underfitting, characterized by unacceptably high $\chi^2$ values ($\sim 2.7-2.9$) and massive systematic deviations (RMSE $> 3000\ \mathrm{km\,s^{-1}}$). Conversely, blindly increasing the structural depth to three layers (C31SN and C32SN) or expanding the width to 8192 neurons (C23SN) leads to performance degradation; the optimization process struggles within the overparameterized space, causing the RMSE and risk to bounce back significantly. However, the dual-layer architecture with 4096 neurons per layer (C22SN) emerges unambiguously as the optimal configuration. It reaches the minimum of the risk curve by simultaneously providing sufficient nonlinear expressiveness to capture the true physical evolution (yielding the lowest RMSE of $1672\ \mathrm{km\,s^{-1}}$) while effectively restraining the exploding variance induced by the correlated noise, hence we eventually reconstruct the $H_0 d_L$ with this configuration. This striking contrast with the CC dataset evaluation underscores a crucial conclusion: the optimal ANN topology is not universal but highly contingent upon the density and covariance complexity of the specific cosmological dataset.

\begin{table}[htbp]
  \centering
  \caption{Configuration notation, parameters, and risk evaluation benchmarks for the extended ANN architectures on the mock SN Ia dataset. The details of each column are as follows: 
  Column 1 (Notation): the specific network configuration; 
  Column 2 ($N$): the number of mock SN Ia data points; 
  Column 3 ($\chi^2$): the reduced chi-square computed with the full covariance matrix; 
  Column 4 (RMSE): the root mean square error representing the average systematic bias, in units of $\mathrm{km\,s^{-1}}$; 
  Column 5 (Var.): the sum of the variance $\sum \mathcal{V}$, in units of $10^{12}\,\mathrm{km^2\,s^{-2}}$; 
  Column 6 (Risk): the total risk balancing bias and variance, in units of $10^{12}\,\mathrm{km^2\,s^{-2}}$.}
  \begin{tabular}{lcccccc}
    \hline\hline
    Notation & $N$ & $\chi^2$ & RMSE & Var. & Risk \\
    \midrule
    C11SN & 1347 & 2.93 & 4430 & 0.258 & 0.289 \\
    C12SN & 1347 & 2.71 & 3085 & 0.258 & 0.273 \\
    \midrule
    C21SN & 1347 & 1.29 & 1767 & 0.255 & 0.260 \\
    C22SN & 1347 & 1.35 & 1672 & 0.254 & 0.259 \\
    C23SN & 1347 & 1.57 & 3338 & 0.259 & 0.277 \\
    \midrule
    C31SN & 1347 & 1.47 & 3872 & 0.253 & 0.277 \\
    C32SN & 1347 & 1.45 & 4508 & 0.256 & 0.288 \\
    \hline\hline
  \end{tabular}
  \label{tab:sn_risk}
\end{table}

\subsection{Long Short-term Memory Networks}\label{Sec:2-LSTM}

For sequential data, LSTM networks are a type of recurrent neural network particularly effective in identifying long-term dependencies and sequential patterns within data. Unlike traditional ANNs, LSTMs' ability to capture these ordered correlations makes them highly suitable for cosmological applications, especially when dealing with redshift-ordered datasets.

To ensure a strictly model-independent reconstruction paradigm, we adopt and structurally modify the LSTM architectures inspired by recent literature \citep{shah2024ladder, cui2025advancing}. In their original implementations, such networks often culminate in a parameter-estimation head that outputs specific cosmological parameters (e.g., $\Omega_m, H_0$). As pointed out in recent methodological discussions, this structure inherently introduces model dependence, as it relies on an assumed background cosmology (e.g., flat $\Lambda$CDM) during the training phase. To rectify this and guarantee a purely data-driven approach, we amputate the original parameter-estimation layer and repurpose the deep LSTM layers strictly as sequential feature extractors. Consequently, rather than predicting physical parameters, our adapted network directly outputs the reconstructed function mean and its corresponding uncertainty (e.g., $H(z)$ and $\sigma_H$) at any target redshift, perfectly aligning with the nonparametric philosophy of our GPR and ANN approaches.

Specifically, to process the sparse CC data ($N=32$), the adapted LSTM is designed as a lightweight architecture containing a single recurrent layer with 16 hidden neurons to effectively prevent overfitting. The training objective is completely shifted from a traditional parameter-fitting metric to a Gaussian Negative Log-Likelihood (NLL) loss. To effectively balance the gradient weights between the highly precise low-redshift observations and the heavily uncertain high-redshift data, an explicit log-space variance supervision is incorporated into the objective. This prevents variance collapse and forces the LSTM to concurrently optimize both the mean trend and its physical error envelope purely from the data's statistical distribution.

Furthermore, to maintain rigorous methodological consistency with our ANN approach, the training process of the LSTM abstains from any early-stopping mechanisms based on validation sets, which could inadvertently introduce data-splitting bias on sparse samples. After strict Z-score normalization of the input data, the network is trained for a fixed length of 1,000 epochs utilizing the Adam optimizer with gradient clipping. An exponential learning rate decay scheme (from $10^{-3}$ down to $10^{-7}$) is employed. This rigorous optimization trajectory ensures that the loss smoothly converges to a stable physical envelope without memorizing the noise. For the much larger SN Ia dataset ($N=1347$), we maintain the original LADDER framework \citep{shah2024ladder}, which is intrinsically designed to reconstruct continuous distance moduli rather than physical parameters, maintaining the model-independent consistency.

\subsection{Nonparametric Reconstructions of Observables}\label{Sec:3-Hz-H0DL}

\begin{figure*}[tbp]
    \centering
    \includegraphics[width=0.7\linewidth]{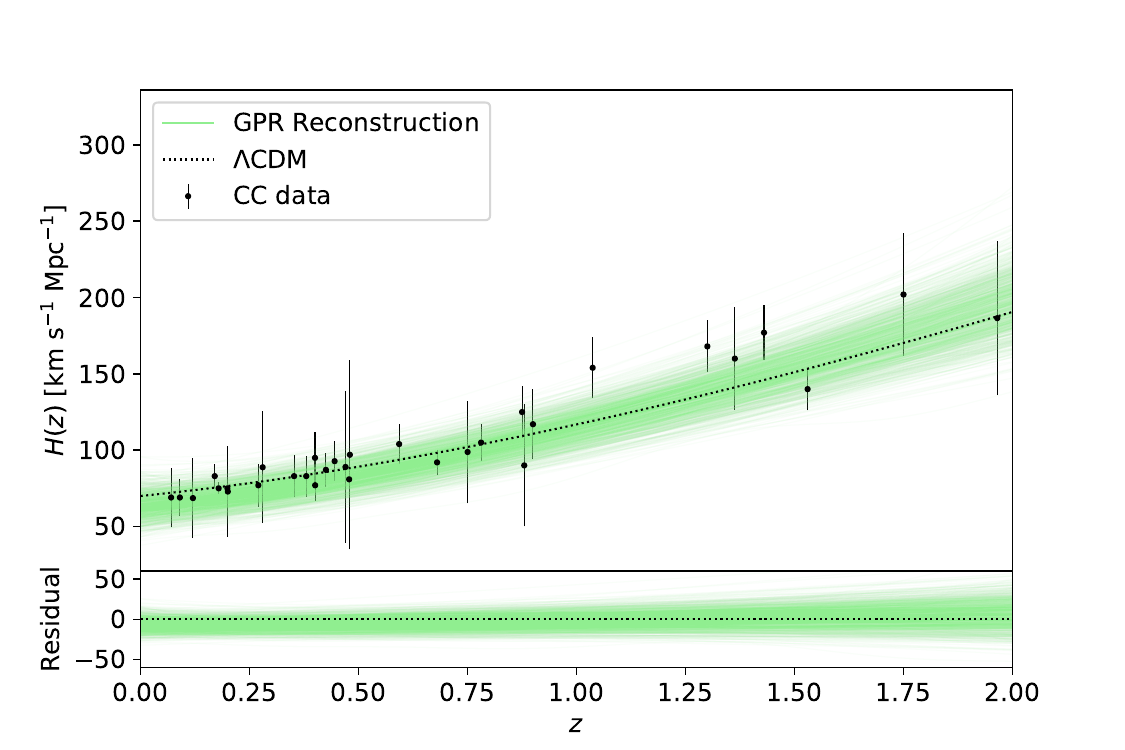}
    \caption{\ttm{The reconstruction of $H(z)$ from the CC dataset based on GPR. \textit{Top panel}: the black points with error bars denote the CC measurements. The light green lines represent 1000 random samples drawn faithfully from the predictive multivariate Gaussian posterior. The black dotted line indicates the $\Lambda$CDM mean function computed using the median values from the MCMC posterior chains. \textit{Bottom panel}: residuals of the 1000 GPR posterior samples with respect to the median $\Lambda$CDM mean function.}}
    \label{fig:Hz_GPR}
\end{figure*}

\begin{figure*}[tbp]
    \centering
    \includegraphics[width=0.65\linewidth]{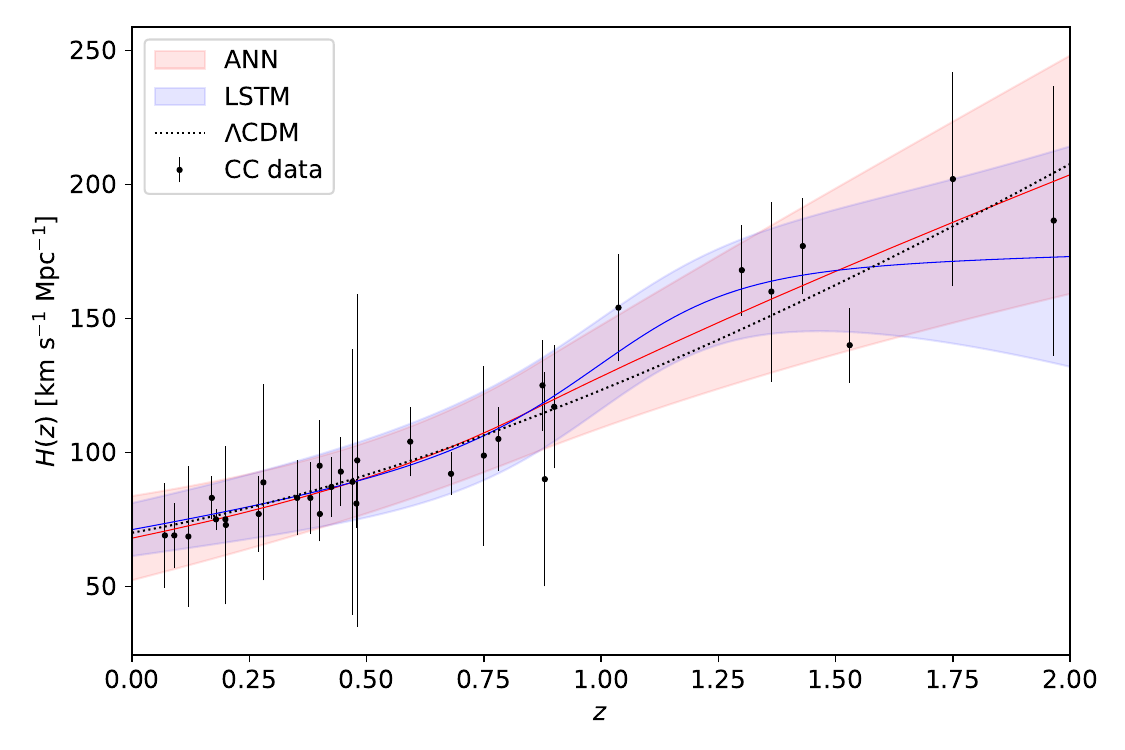}
    \caption{The reconstruction of $H(z)$ from the CC dataset based on ANN and LSTM. The red line and shadow show the ANN's reconstructed mean values and $1\sigma$ uncertainties, while the blue line and shadow represent the LSTM's predictions. The $H(z)$ from the fiducial flat $\Lambda$CDM model is shown as the black dotted line.}
    \label{fig:Hz_ANN_LSTM}
\end{figure*}

Following the methodologies detailed in Sections~\ref{Sec:2-GPR}--\ref{Sec:2-LSTM}, we first apply the GPR, ANN, and LSTM frameworks to reconstruct the smooth, continuous expansion history $H(z)$ from the CC dataset. For the GPR approach, the fully marginalized MCMC sampling ensures that the reconstruction is unanchored from the fiducial mean function, faithfully capturing the full posterior distribution of $H(z)$. \ttm{In addition, by incorporating the full covariance matrix (using \texttt{imf}, \texttt{slib}, and \texttt{spsooo}), this reconstruction efficiently avoids the underestimation of uncertainties. The resulting GPR reconstruction and its predictive multivariate Gaussian samples are shown in Figure~\ref{fig:Hz_GPR}.} In parallel, the purely data-driven ANN and LSTM networks are trained on the CC data to learn the direct mapping from redshift to the Hubble parameter. As established in our methodology, the ANN minimizes the mean absolute error, while the LSTM optimizes a Gaussian NLL loss with log-space variance supervision. Both machine learning approaches successfully provide smooth, physically consistent $H(z)$ functions without assuming any background cosmology, as displayed in Figure~\ref{fig:Hz_ANN_LSTM}.

Focusing next on the Pantheon+ SN Ia sample, we reconstruct the unanchored luminosity distance $H_0 d_L(z)$. To ensure our results are completely independent of local distance ladder calibrations, we strictly select the ``model-independent'' samples by filtering for \texttt{IS\_CALIBRATOR = 0} and \texttt{USED\_IN\_SH0ES\_HF = 0}. Furthermore, we implement a rigorous statistical treatment to incorporate the full systematic and statistical covariance matrix into the GPR framework. Since the observational covariance matrix $\mathbf{C}_{m}$ is provided in terms of the apparent magnitude $m_B$, while our reconstructed state variable is $H_0 d_L$, we perform a nonlinear error propagation using a Jacobian transformation:
\begin{equation}
    \mathbf{C}_{H_0 d_L} = \mathbf{J} \mathbf{C}_{m} \mathbf{J}^T + \mathbf{C}_{\text{calib}},
\end{equation}
where the Jacobian elements are defined as $\partial (H_0 d_L) / \partial m_B = (H_0 d_L) \ln(10) / 5$. Additionally, we explicitly account for the global calibration uncertainty of the absolute magnitude by adding a fully correlated covariance term $\mathbf{C}_{\text{calib}}$, ensuring that the systematic shift in the absolute scale is accurately propagated through the sampling process.

\begin{figure*}[tbp]
    \centering
    \includegraphics[width=0.7\linewidth]{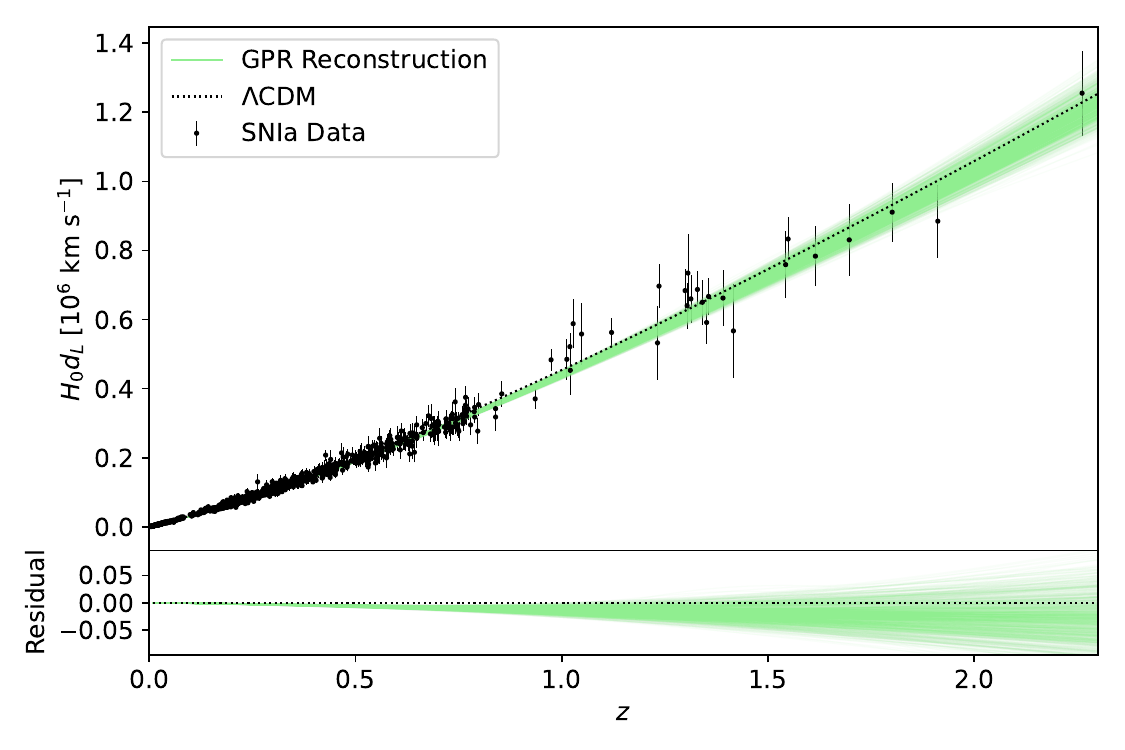}
    \includegraphics[width=0.68\linewidth]{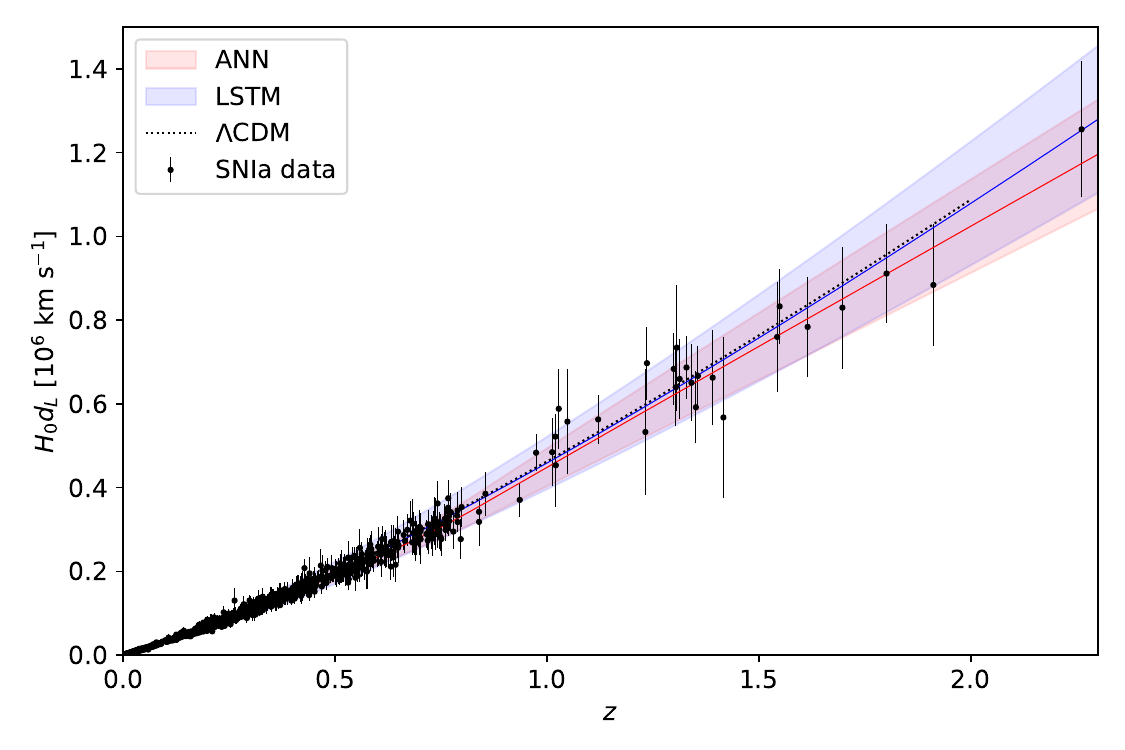}
   \caption{\ttm{The nonparametric reconstruction of $H_0d_L(z)$ from Pantheon SNe Ia, based on three machine learning methods.}}
  \label{fig:H0DL_rec}
\end{figure*}

For the machine learning techniques applied to SN Ia, the ANN is trained directly to map the redshifts to $H_0 d_L$, while the LSTM utilizes the LADDER framework to handle the complex covariance information robustly. As displayed in Figure~\ref{fig:H0DL_rec}, all three nonparametric techniques yield consistent, smooth reconstruction curves that successfully capture the intrinsic trends of the unanchored SN Ia distances while faithfully preserving the observational uncertainties.

\begin{table}[htbp]
    \centering
    \caption{\ttm{Model-independent constraints on the Hubble constant $H_0$ using DESI DR2 BAO combined with SN Ia and CC data reconstructed via GPR, ANN, and LSTM.}}
    \renewcommand\arraystretch{1.2}
    \begin{tabular}{lllll}
        \hline\hline
        \multicolumn{2}{l}{Reconstructor} & Tracer & $H_0$\\
        SN Ia & CC & & $\mathrm{km\,s^{-1}\,Mpc^{-1}}$\\
        \hline
        GPR & GPR & LRG1 & $58.4\pm\,\,\, 5.8$ \\
         & & LRG2 & $60.5\pm\,\,\, 5.7$ \\
         & & LRG3~+~ELG1 & $63.3\pm\,\,\, 5.5$ \\
         & & ELG2 & $63.8\pm\,\,\, 5.5$ \\
         & & QSO & $61.6\pm\,\,\, 6.3$ \\
         & & Ly$\alpha$ & $63.0\pm\,\,\, 6.7$ \\
         & & \textbf{Joint} & $61.8\pm\,\,\, 5.3$ \\
        \hline
        ANN & ANN & LRG1 & $63.0\pm 10.4$ \\
         & & LRG2 & $64.6\pm 10.5$ \\
         & & LRG3~+~ELG1 & $70.9\pm 12.5$ \\
         & & ELG2 & $72.0\pm 14.7$ \\
         & & QSO & $68.4\pm 15.1$ \\
         & & Ly$\alpha$ & $61.0\pm 16.0$ \\
         & & \textbf{Joint} & $66.3\pm\,\,\, 5.2$ \\
        \hline
        LSTM & LSTM & LRG1 & $65.0\pm 13.9$ \\
         & & LRG2 & $67.3\pm 14.3$ \\
         & & LRG3~+~ELG1 & $74.4\pm 14.5$ \\
         & & ELG2 & $77.4\pm 14.1$ \\
         & & QSO & $70.6\pm 14.1$ \\
         & & Ly$\alpha$ & $50.3\pm 18.3$ \\
         & & \textbf{Joint} & $68.7\pm\,\,\, 6.0$ \\
        \hline\hline
    \end{tabular}
    \label{tab:H0_results}
\end{table}

\begin{table}[htbp]
    \centering
    \caption{\ttm{Joint constraints on the Hubble constant $H_0$ by systematically excluding individual DESI BAO tracers. The ``Baseline'' row represents the fiducial joint constraints derived using the complete set of all six tracers, provided for direct comparison.}}
    \vspace*{1.5em}
    \renewcommand\arraystretch{1.2}
    \begin{tabular}{lccc}
        \hline\hline
        Excluded & \multicolumn{3}{c}{$H_0$ ($\mathrm{km\,s^{-1}\,Mpc^{-1}}$)} \\
        \cline{2-4}
        Tracer & GPR & ANN & LSTM \\
        \hline
        LRG1       & $62.5 \pm 5.3$ & $67.3 \pm 6.0$ & $69.5 \pm 6.7$ \\
        LRG2       & $62.1 \pm 5.3$ & $66.8 \pm 5.9$ & $69.0 \pm 6.6$ \\
        LRG3+ELG1  & $61.4 \pm 5.2$ & $65.3 \pm 5.7$ & $67.5 \pm 6.6$ \\
        ELG2       & $61.3 \pm 5.3$ & $65.5 \pm 5.5$ & $66.8 \pm 6.6$ \\
        QSO        & $61.8 \pm 5.3$ & $66.0 \pm 5.5$ & $68.3 \pm 6.6$ \\
        Ly$\alpha$ & $61.6 \pm 5.3$ & $66.9 \pm 5.5$ & $70.9 \pm 6.4$ \\
        \hline
        \textbf{Baseline} & $61.8 \pm 5.3$ & $66.3 \pm 5.2$ & $68.7 \pm 6.0$ \\
        \hline\hline
    \end{tabular}
    \label{tab:excluded_data}
\end{table}

\begin{figure*}[tbp]
    \centering
    \includegraphics[width=0.7\linewidth]{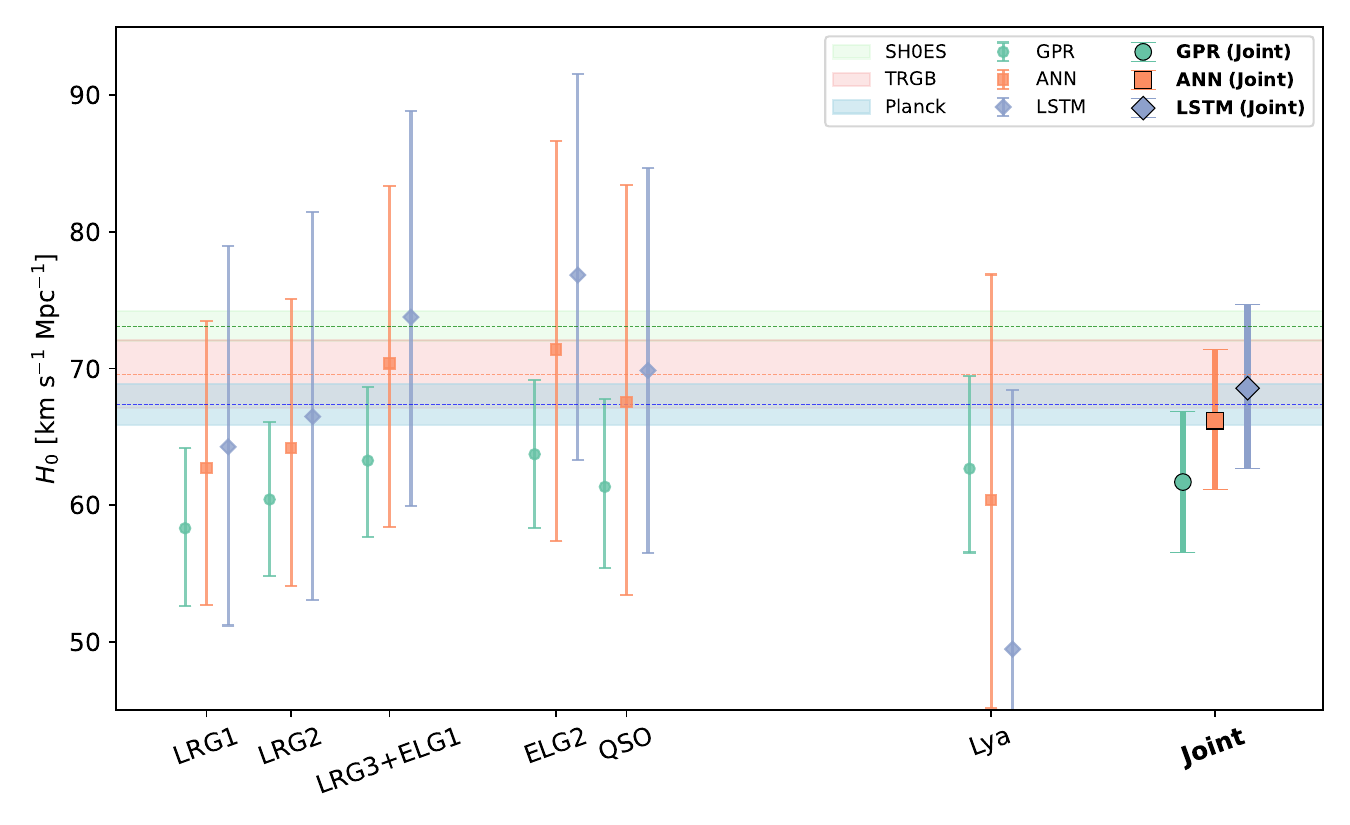}
    \caption{\ttm{Posterior distributions of the Hubble constant $H_0$ derived from DESI DR2 BAO and reconstructed CC+SN Ia data, based on GPR (green), ANN (orange), and LSTM (blue). Vertical bands show the $1\sigma$ regions of SH0ES \citep{Pantheon2022}, TRGB \citep{TRGB2020}, and Planck \citep{Planck2020}.}}
    \label{fig:H0_PDF}
\end{figure*}

\begin{figure*}[tbp]
    \centering
    \includegraphics[width=0.7\linewidth]{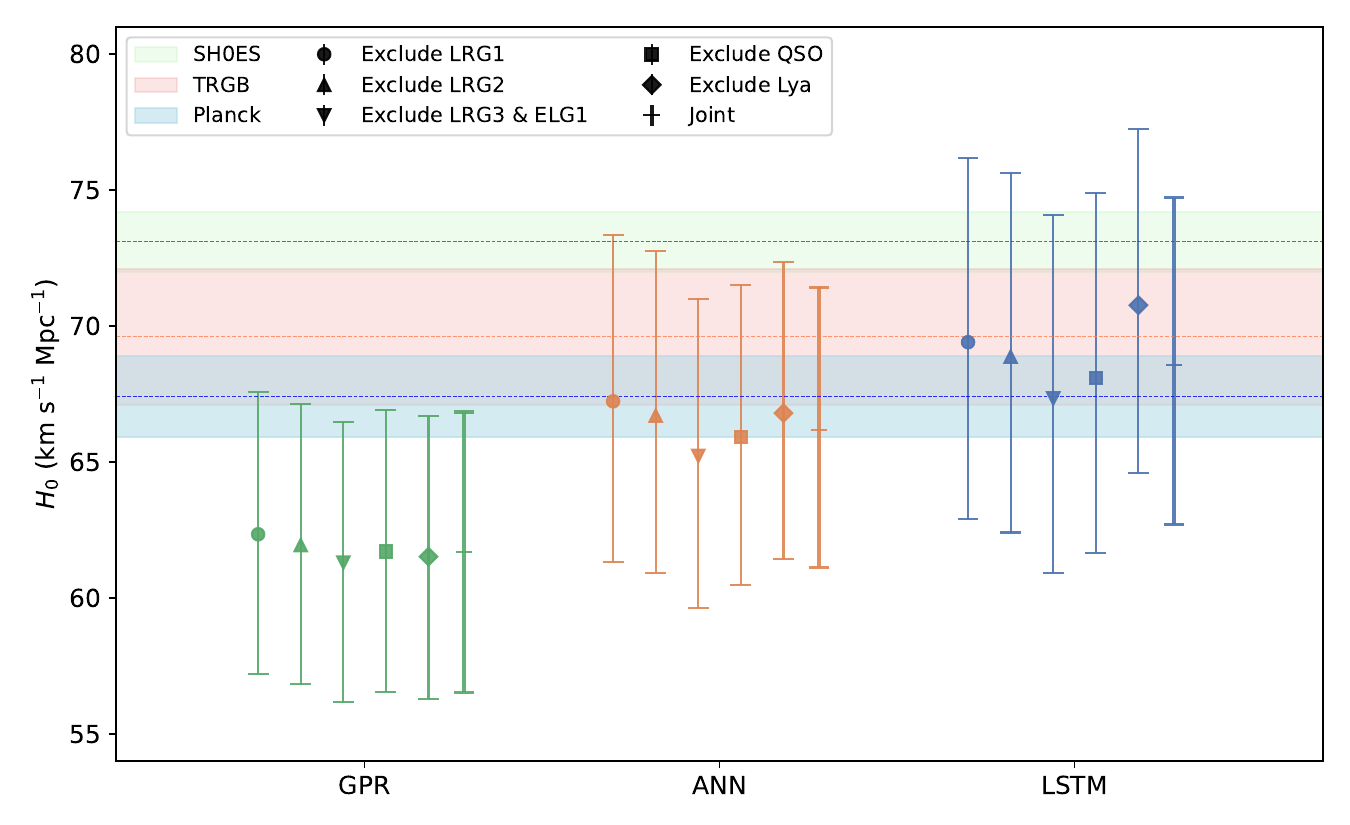}
    \caption{\ttm{Impact of the individual DESI BAO data point on the inferred Hubble constant $H_0$. Results show that the removal of any single tracer causes a significant shift in $H_0$ at the $1-2\ \mathrm{km\ s^{-1}\ Mpc^{-1}}$ level. 
    }}
    \label{fig:H0_sensitivity}
\end{figure*}

\section{Results and Discussion}\label{Sec:4}

With these nonparametric reconstructions applied to $\Xi_\text{SN Ia}(z)$ and $H_\text{CC}(z)$, we can obtain the $H_0$ measurement given $\Xi_\text{BAO}(z)$ at each BAO redshift (Equation~(2)), the uncertainty of which is directly propagated from the posteriors of reconstructions. Based on the observable $D_M(z)/D_H(z)$ at BAO redshift $z$, one can derive $\Xi_\mathrm{BAO}(z) = H(z)d_A(z)$ from Equation~(\ref{Eq:HzdA}). Since the anisotropic parameter $\alpha_\mathrm{AP}$ is directly estimated from the fitting-to-clustering data \citep{adame2025desi}, instead of using the $D_M/D_H$ given by Table IV from \cite{DESI2025}, we utilize $\alpha_\mathrm{AP}$ and calculate 
\begin{equation}
    \frac{D_M(z)}{D_H(z)} = \frac{1}{\alpha_\mathrm{AP}}\frac{D_M^\mathrm{fid,Planck}(z)}{D_H^\mathrm{fid,Planck}(z)}.\label{Eq:DMDH}
\end{equation}
Note that $D_M^\mathrm{fid,Planck}(z)/D_H^\mathrm{fid,Planck}(z)$ is calculated from a flat $\Lambda$CDM model with \cite{Planck2020} results, which is used throughout the DESI DR2 data processing. Then the MC sampling is performed to $\alpha_\mathrm{AP}$ to derive $\Xi_\mathrm{BAO}$ with Equation~(\ref{Eq:HzdA}) and (\ref{Eq:DMDH}), so that one can capture the non-Gaussian distribution of $\Xi_\mathrm{BAO}$ at each BAO point.

For ANN and LSTM, the networks directly predict the mean values and uncertainties at each BAO redshift, from which MC samples are drawn. For the GPR method, we strictly adhere to a Bayesian framework. Because the direct output of GPR is already a multivariate Gaussian distribution, fitting independent Gaussian distributions at each redshift would incorrectly discard the cross-redshift correlations. Instead, we analytically construct the full predictive covariance matrix $\text{cov}(f^*)$ at the targeted BAO redshifts. The joint multivariate Gaussian distribution $\mathcal{N}(\bar{f}^*, \text{cov}(f^*))$ naturally accounts for the spatial dependencies inherent in the cosmic expansion history. To derive the posterior of $H_0$, we employ a MC strategy by randomly drawing joint samples from the reconstructed multivariate distributions of $H_\mathrm{CC}(z)$ and $\Xi_\mathrm{SN Ia}(z)$, as well as the observational $\Xi_\mathrm{BAO}(z)$. This MC approach is fully sufficient and entirely avoids the unnecessary computational burden of a full grid-like Cartesian combination, while guaranteeing that the cross-redshift correlations between different BAO points are faithfully preserved. For the final joint constraint on $H_0$, an MCMC method is implemented with a multivariate Gaussian likelihood. The individual and joint constraint results are summarized in Table~\ref{tab:H0_results} and Figure~\ref{fig:H0_PDF}.

The three distinct reconstruction strategies---GPR, ANN, and LSTM applied to both CC and SN Ia datasets---produce statistically consistent estimates of $H_0$ at individual BAO points, as well as comprehensive joint constraints on $H_0$. \ttm{The GPR method yields a constraint of $H_0 = 61.8 \pm 5.3\ \mathrm{km\,s^{-1}\,Mpc^{-1}}$ by explicitly modeling the fully correlated systematic covariance structure. The purely data-driven ANN and LSTM approaches yield $H_0 = 66.3 \pm 5.2\ \mathrm{km\,s^{-1}\,Mpc^{-1}}$ and $H_0 = 68.7 \pm 6.0\ \mathrm{km\,s^{-1}\,Mpc^{-1}}$, which accurately reflects their highly nonlinear flexibility and the lack of strong structural priors. The consistent central values derived from these diverse machine learning architectures confirm the robustness of our overall reconstruction methodology. Our model-independent constraints exhibit excellent agreements with the early-Universe inference at $1\sigma$ level \citep{Planck2020}. Furthermore, the ANN and LSTM results also agree well with recent geometric constraints \citep{Renzi2023} and TRGB observations \citep{TRGB2020} within $1\sigma$. Our method avoids calibrating the absolute magnitude of SN Ia, thereby differing from local distance ladder approaches. Notably, the LSTM constraint is marginally consistent with that of the SH0ES collaboration, which reports a significantly higher value of $H_0 = 73.04 \pm 1.04~\mathrm{km\,s^{-1}\,Mpc^{-1}}$ based on Cepheid-calibrated supernovae \citep{riess2022comprehensive}. }

We further investigate the effect of dataset redshift coverage and potential systematics. Among the six DESI BAO points, the Ly$\alpha$ point at $z = 2.33$ lies beyond the maximum redshift of the CC data ($z \leq 1.965$). Since its inclusion requires extrapolation in nonparametric reconstructions, we assess the outcome of excluding it from the analysis. \ttm{The resulting joint constraints are $H_0^\mathrm{GPR;-\text{Ly}\alpha} = 61.6\pm 5.3~\mathrm{km\,s^{-1}\,Mpc^{-1}}$, $H_0^{\mathrm{ANN;-\text{Ly}\alpha}} = 66.9\pm 5.5~\mathrm{km\,s^{-1}\,Mpc^{-1}}$, and $H_0^{\mathrm{LSTM;-\text{Ly}\alpha}} = 70.9\pm 6.4~\mathrm{km\,s^{-1}\,Mpc^{-1}}$. While removing the extrapolated Ly$\alpha$ point slightly shifts the ANN and LSTM central values upward, these results remain highly consistent with the baseline results in Table~\ref{tab:H0_results}, indicating that the extrapolation has minimal impact.} In order to test the sensitivity to specific DESI tracers---particularly given the mild tension reported between DESI LRG measurements and earlier SDSS results \citep{DESI2025,WangHu2024}---we perform a comprehensive leave-one-out analysis for all BAO points. \ttm{As summarized in Figure~\ref{fig:H0_sensitivity} and Table~\ref{tab:excluded_data}, systematically excluding individual tracers yields shifts in the inferred $H_0$ central values of merely $\lesssim 2.5\,\mathrm{km\,s^{-1}\,Mpc^{-1}}$ across different methods}. Crucially, all these shifts remain well within the $1\sigma$ statistical uncertainties of the baseline constraints, confirming that our joint results are not unduly biased by any single measurement.

In summary, this paper provides a robust, purely data-driven constraint on the Hubble constant using the latest DESI DR2 BAO measurements combined with nonparametrically reconstructed SN Ia and CC data. The internal consistency among GPR, ANN, and LSTM, combined with their stability against specific BAO exclusions and their agreement with early-Universe probes, solidly validates our methodological approach. The fact that our low- and intermediate-redshift geometric constraints prefer a lower $H_0$ implies that the tension cannot be simply explained by late-time modifications to the expansion history. Such a conclusion is also drawn from extensive parametric studies \citep{keeley2023ruling}. \ttm{Furthermore, our nonparametric frameworks demonstrate that incorporating comprehensive, model-independent uncertainties (such as the fully correlated CC covariance) can yield broad statistical compatibility with both early-Universe predictions and local distance ladder measurements.} Future improvements in CC observations and a deeper understanding of systematic errors in DESI tracers will further refine this framework and contribute to resolving the Hubble tension.

\section*{Acknowledgments}

This work is supported by Beijing Natural Science Foundation No. 1242021, the National Natural Science Foundation of China under grant No. 12433001, and the Fundamental Research Funds for the Central Universities. T.L. was supported by the National Natural Science Foundation of China under grant No. 12203009, Chutian Scholars Program in Hubei Province (X2023007), and Hubei Province Foreign Expert Project (2023DJC040). M.B. was supported by the Polish National Science Centre grant 2023/50/A/ST9/00579 and by COST Action CA21136---``Addressing observational tensions in cosmology with systematics and fundamental physics (CosmoVerse).''

\bibliography{reference.bib}{} 
\bibliographystyle{aasjournal}

\end{document}